\documentclass[11pt]{article}
\usepackage{moriond,epsfig}
\usepackage{graphicx}
\def\Journal#1#2#3#4{{\em #1} {\bf #2}, #3 (#4)}

\def\be{\begin{equation}}
\def\ee{\end{equation}}
\def\bc{\begin{center}}
\def\ec{\end{center}}
\def\bea{\begin{eqnarray}}
\def\eea{\end{eqnarray}}

\def\msun{{M_\odot}}
\def\xte{{\it RXTE}}
\def\deg{^{\circ}}
\newcommand\fdg{\mbox{$.\!\!^\circ$}}
\def\rg{r_g}
\def\sax{SAX J1808.4-3658}

\begin{document}
\vspace*{4cm}
\title{MODELING THE ENERGY DEPENDENT PULSE PROFILES OF THE ACCRETING MILLISECOND 
PULSAR SAX J1808.4-3658}

\author{Juri Poutanen$^{1}$ and Marek Gierli\'nski$^{2,3}$}

\address{$^1$Astronomy Division, P.O. Box 3000, FIN-90014 University of Oulu,
Finland \\
$^2$ Department of Physics, University of Durham, South Road, Durham DH1 3LE, UK\\
$^3$ Astronomical Observatory, Jagiellonian University, Orla 171, 30-244 Krak\'ow,
Poland}

\maketitle

\abstracts{The pulse profiles  of the accreting X-ray millisecond  pulsar
\sax\ at different energies   are studied.
The two main emission component, a black body and a power-law tail,
clearly identified in the time-averaged spectrum, do not vary in phase.
We show that the observed variability can be easily explained if the
emission patterns of the black body and the Comptonized radiation
are different: a ``knife" and a ``fan"-like, respectively.
We suggest that Comptonization in  a hot slab (radiative shock) of
Thomson optical depth $\sim$ 0.3
at the surface of the neutron star may be responsible for the emission.
We construct a detailed model of the X-ray production
accounting  for the Doppler boosting, relativistic  aberration and
gravitational light bending. The model reproduces well
the  pulse profiles at different energies simultaneously, corresponding phase lags,
as well as the time-averaged spectrum. 
By fitting the observed pulse profiles we obtain constraints on the neutron 
star radius ($R=7.5\pm1.0$ km), the inclination of the system $i>60\deg$, 
and the angle between the magnetic dipole and the rotational axis 
$\delta=20\deg\pm 5\deg$. 
}

\section{Introduction}

The X-ray burster \sax\ was detected in September 1996
using the Wide Field Camera aboard {\it BeppoSAX} (in't Zand et al. 1998).
Coherent pulsation with a period of 2.49 ms were detected
in the {\it Rossi X-ray Timing Explorer} data of the April 1998 outburst
of this source (Wijnands \& van der Klis 1998).
Chakrabarty \& Morgan (1998)  have discovered changes in the pulse
arrival corresponding to the binary period of 2 hours.
During the 1998 outburst, the source energy spectrum
did not change much and the coherent pulsations with rms amplitude of 5-7 per cent
were detected in all but one observation (Gilfanov et al. 1998, Cui et al. 1998)
in spite of the 100-fold decrease of the luminosity.
The pulse profiles show strong energy dependence and soft phase lags (Cui et al. 1998).
By analyzing the  phase-resolved spectra,
Gierli\'nski et al. (2002) showed that the two main spectral components
(soft black body and hard Comptonized) do not vary in phase.
The black body emission is lagging the Comptonized emission.
This causes changes in the pulse profile with energy and produces the
observed soft phase  lags. While the black body light curve is nearly sinusoidal,
the hard component is clearly asymmetric and cannot be fitted by a
single harmonic. Gierli\'nski et al. suggested that the Doppler effect can
play a role in changing the pulse profile at high energies.

The source was observed 12 times by the \xte\ during the April 1998 outburst.
We construct the energy-dependent pulse profiles following the
procedure described in Gierli\'nski et al. (2002) correcting the photon 
arrival times for orbital motions of the pulsar and the spacecraft. 
In order to improve the statistics, we average the profiles
using the data from 11--26 April 1998.

In this paper, we construct a detailed model for the X-ray emission from
a hot spot at the surface of the neutron star accounting for
relativistic effects. We compare the model with the data and
constrain the model parameters.

\section{Model}

\subsection{Incident spectrum}

\begin{figure}\bc
\psfig{figure=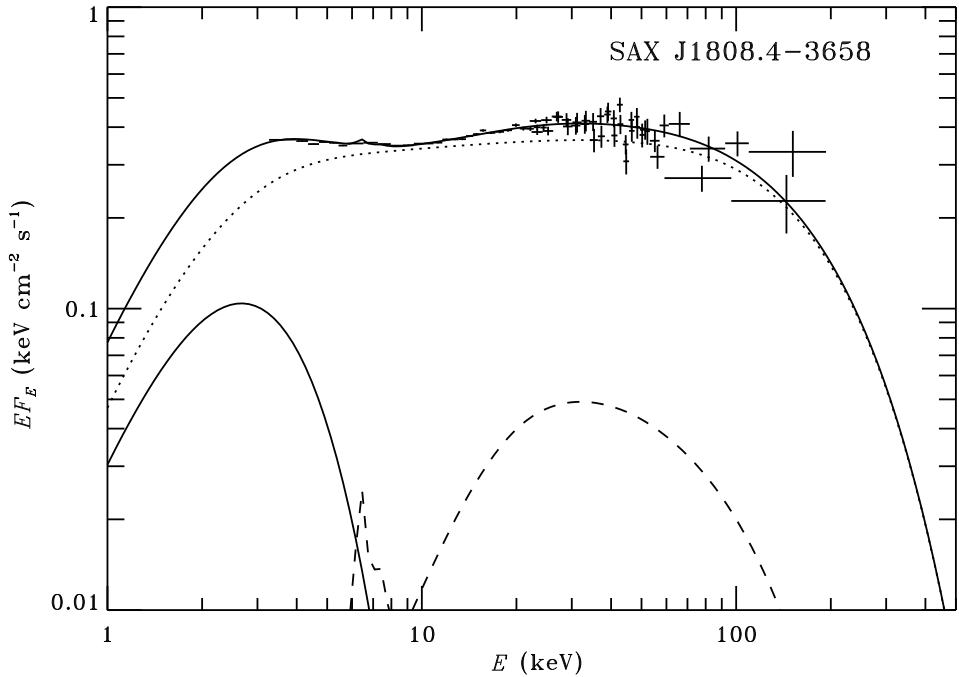,height=3.5in}\ec
\caption{Time-averaged spectrum of \sax\ as observed
by \xte. The model spectrum  (upper solid curve)
consists of a black body (lower solid curve),
thermal Comptonization (dotted curve),
and Compton reflection with the iron line (dashed curve).
Thermal Comptonization component (model {\sc thcomp} in XSPEC; Zdziarski, Johnson,
\& Magdziarz 1996) has a best-fit photon spectral slope $\Gamma=1.92$.
The electron temperature was fixed at 90 keV and the black body temperature
at $T_{\rm bb}=0.68$ keV.
\label{fig:speave}}
\end{figure}

The time-averaged spectrum of SAX J1808.4-3658 can be decomposed into a power-law like 
Comptonized component extending at least to 200 keV, a black body (contributing
about 30\% to the flux in the 2--5 keV region) and a weak Compton reflection bump
corresponding to a covering factor $\Omega/(2\pi)\sim 0.1$
with an iron line (see Fig.~\ref{fig:speave} and Gilfanov et al. 1998,
Gierli\'nski et al. 2002). The angular distribution of these 
components is however unknown. 
We assume that the emission originates in a relatively small
spot (or two diametrically opposite spots)  at the magnetic pole of a neutron star. 
In the frame of the spot, the angular distribution of radiation is azimuthally 
symmetric and depends only on the zenith angle $\theta$.

\begin{figure}
\bc
\psfig{figure=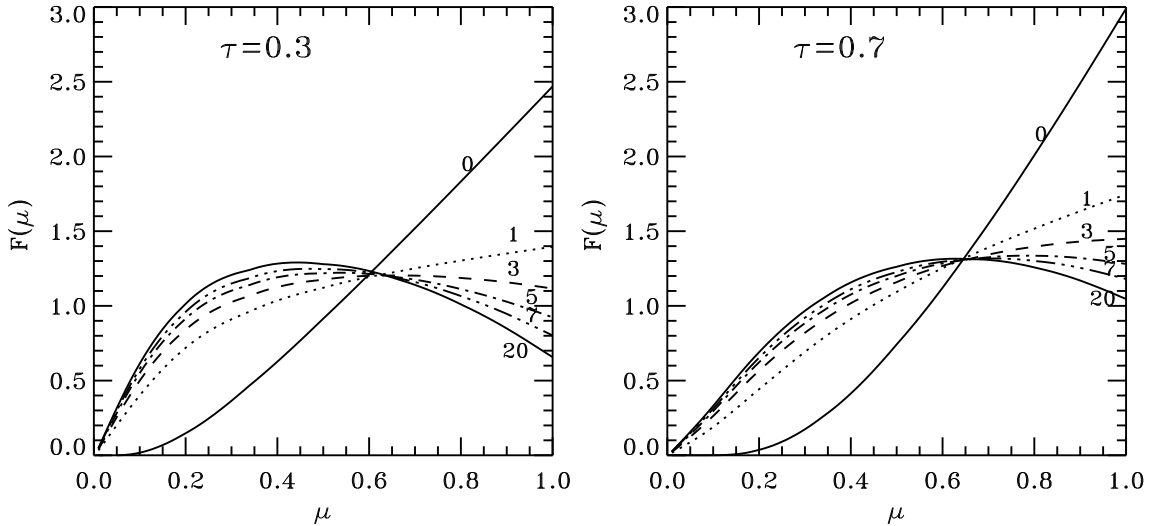,width=6in}
\ec
\caption{Angular distribution of the radiation flux (normalized as $\int F(\mu) d\mu=1$)
escaping from a purely scattering 
slab of Thomson optical depth $\tau=0.3$ ({\it left}) and $\tau=0.7$ ({\it right}). 
Incident radiation from the bottom is a black body (intensity does not depend 
on the zenith angle $\arccos \mu$). 
Computations follow procedure described in Sunyaev \& Titarchuk (1985).
Polarization is accounted for. 
Different scattering orders are shown and marked by numbers. 
In the Comptonization process, scatterings also shift photons along the
energy axis so that at a given energy one scattering order dominates
(see e.g. Poutanen \& Svensson 1996).
\label{fig:ray}}
\end{figure}

For a slab-like geometry of the hot emitting region,
the black body flux entering that region from the bottom 
can be represented as $F_{\rm bb}(\mu)\propto \mu \exp(-\tau/\mu)$,
where $\mu=\cos\theta$ and $\tau$ is the Thomson optical thickness
of the Comptonizing layer.
In contrast to the black body radiation strongly peaking
in the normal direction, the scattered radiation is expected to have a much broader
distribution peaking not necessarily along the normal 
(see Fig.~\ref{fig:ray} and Sunyaev \& Titarchuk 1985).
One should remind the reader that the photons increase their energy with
each subsequent scattering. Therefore, at energies far above
the black body peak, the spectrum is dominated by scattered photons.
After a few scatterings the angular distribution does not change anymore.

In principle, $\tau$ can be obtained by fitting the broad-band X-ray data.
However, because of the low signal above 100 keV, the electron temperature $T_e$
is not well constrained. Fixing it at 90 keV and fitting the data by a
Comptonization model {\sc compps} in the slab geometry (Poutanen \& Svensson 1996),
we obtain $\tau=0.5$. At lower temperatures, the best-fit optical depth is higher,
e.g. for $kT_e=60$ keV, $\tau=0.9$. Thus, we just  parametrize the angular distribution 
of scattered photons  as $F_{\rm sc}(\mu) \propto \mu (1+b\mu)$, where $b$ is 
a parameter depending on the optical depth $\tau$ and the scattering order.

\subsection{Model light curves}

We assume that in the rest frame of the spot the black body
and the Comptonized emission have the same energy dependence
as the time-averaged spectra (see Fig.~\ref{fig:speave}).
We neglect the gravitational redshift since it acts in exactly
the same way to photons  of all energies.
Assuming different emission pattern for the black body and
the Comptonized emission we proceed to computing the expected pulse
profiles at different energies. For simplicity, we assume that 
the angular distribution of the Comptonized radiation does not 
depend on energy. 
We account for the
special relativistic effects (Doppler boosting, relativistic
aberration), and light bending in Schwarzschild geometry (Pechenick et al. 1983,
Beloborodov 2002). The computed black body and Comptonized components are then
renormalized  to their best-fit time-averaged values.

The parameters of the model are the neutron star mass $M$, its radius $R$,
the rotational frequency, $\nu$ (401 Hz for \sax), the inclination $i$,
the angle $\delta$ between the magnetic dipole and the rotational axis,
and two parameters describing the angular distribution of the
incident radiation $\tau$ and $b$.

Since the observed profiles have almost sinusoidal shapes (at least at lower energies),
we  compute the light curves from the primary spot (closest to the observer) only.
For the same reason, we neglect multiple light revolutions around a neutron star 
which are possible for $R<1.5 \rg$ (here $\rg\equiv 2GM/c^2$).   
For almost all equations of state, the secondary, antipodal, spot should be visible
due to the light bending effects (see Beloborodov 2002). We therefore
conclude that the light of sight to the secondary spot is most probably blocked by
the accretion disk.

\section{Results}

\subsection{Role of the parameters}

Before we proceed with the data fitting, we would like to discuss 
the expected correlations between different parameters. 
For example, one would expect some ambiguity regarding the exchange of the 
angles $i$ and $\delta$. One can easily 
understand that considering a single spot emitting
black body radiation (flux is proportional to $\mu$).
Using an approximate description of the light bending effect
in Schwazschild spacetime (Belorobodov 2002), one can relate
the oscillation amplitude, $A\equiv (F_{\max}-F_{\min})/(F_{\max}+F_{\min})$,
to the angles $i$ and $\delta$ and the neutron star radius:
\be
A=\frac{\sin i\ \sin \delta}{\cos i \ \cos\delta+ 1/(R/\rg-1)}.
\ee
Thus for a given radius there is a curve at the
$i-\delta$ plane producing a given variability amplitude. 
For a nonzero optical depth $\tau$, the black body component
becomes more beamed towards the spot normal thus increasing $A$.

\begin{figure}\bc
\psfig{figure=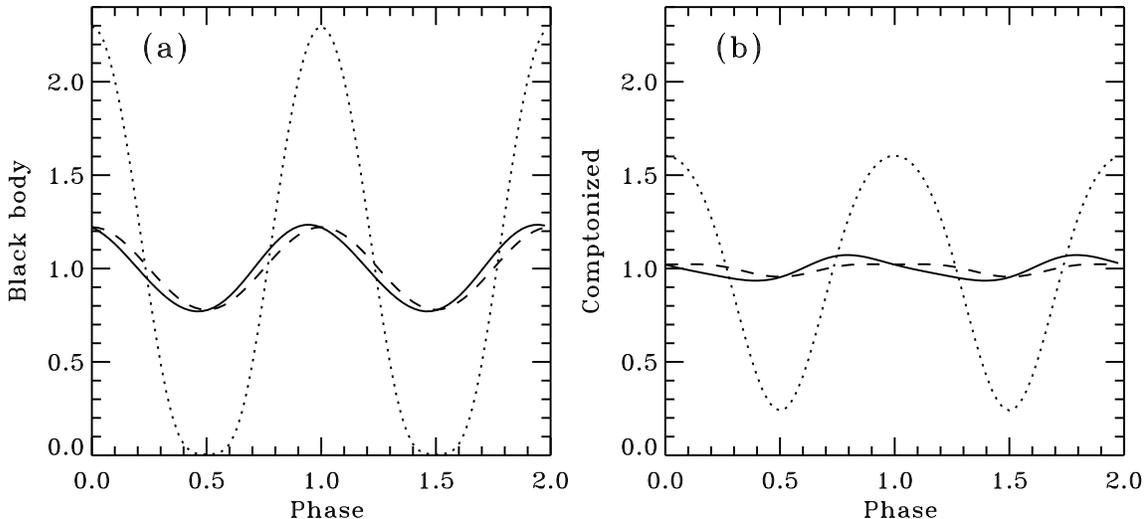,width=6.in}\ec
\caption{(a) The black body flux and  
(b) the Comptonized flux normalized to the mean value.
Dotted curves show  the dependences where no special or general relativistic
effects are taken into account. Dashed curves correspond to the profiles
where  light bending in the Schwarzschild geometry is accounted for. 
The profiles modified by Doppler boosting and aberration as well as light bending 
are shown by solid curves. For the model parameters see Fig.~\ref{fig:all}.
\label{fig:relateff}}
\end{figure}

For a typical neutron star radius of 10 km, the velocity 
at the equator reaches $\sim 0.1c$. Thus one expect that the Doppler effect 
plays an important role significantly distorting the pulse profiles. 
Since the maxima of the Doppler factor and  the spot projected area are shifted in 
phase by $\sim$ 0.25  (phase varies between 0 and 1), the peak in the emission will be shifted 
towards the phase where $D$ has the maximum (i.e. $\phi=0.75$, see Fig.~\ref{fig:relateff}; 
$\phi=0$  corresponds to the position of the spot closest to the observer).
The Doppler factor $D=1/[\gamma(1-\beta\cos\xi)]$, 
where $\beta=v/c=2\pi (R/c)\nu \sin\delta$ is the velocity of the spot,
$\gamma=1/\sqrt{1-\beta^2}$ is the Lorentz factor, and  $\xi$ is the 
angle between the spot velocity vector and the emitted photon momentum.  
Note, that in the absence of light bending, $\cos\xi=-\sin i \sin(2\pi\phi)$. 
Thus again interchanging $i$ and $\delta$ does not affect the Doppler factor 
much. However, the Doppler factor varies slightly along the $A=const$ curve. 
This helps to break the degeneracy between the parameters. 

The absence of the secondary spot emission also implies that
it is blocked by the accretion disk. 
At inclinations smaller that $45\deg$ 
the upper limit on the inner disk radius required to block the secondary spot is 
$R_{\rm in}<2R$. With such a small inner radius, 
the expected amplitude of Compton reflection component is much larger than observed 
$\Omega/(2\pi)\sim 0.1$. 
Large inclination is also consistent with the lower limit
$i> 28\deg$ obtained by Wang et al. (2001)  based on the modeling of the optical/IR
emission with the X-ray irradiated disk. Thus we consider small inclination very 
unprobable.

The radius of the neutron star affects the profiles in two senses. 
A larger radius corresponds to a larger velocity of the spot and thus to a 
larger Doppler factor. 
At the same time, for a larger $R$ the light bending becomes less important
and the amplitude of variability $A$ increases.

\subsection{Data fitting}

We choose to fit the observed pulse profiles in the energy bands 3--4 keV and 
12--18 keV simultaneously.
In the lower energy band, both emission components play an important role, 
while the second band is dominated by the Comptonized radiation.

The mass of the neutron star in \sax\ is not know. As an example, 
we consider a mass of $M=1.6\msun$. 
(One would expect $1.5\msun<M<2\msun$ for neutron stars in LMXRBs that have 
accreted material from their binary companion.)   
The 90\% confidence interval (corresponding to $\Delta\chi^2=2.7$) for the neutron star 
radius becomes $R=7.5\pm1.0$ km.
The preferred inclination is rather large $i>60\deg$.
We note that the upper limit on the inclination   $i<83\deg$ 
is derived by  Bildsten \& Chakrabarty (2001) from the absence of the X-ray eclipses. 
Other  model parameters are $\delta=20\deg\pm5\deg$,  $\tau=0.3\pm0.1$, and 
$b=-0.65\pm0.1$ with the minimum $\chi^2=47.4$ for 34 dof. 
The obtained value for $b$ corresponds to the optical depth of about 0.7, 
while the best-fit $\tau$ is much smaller. If the geometry of the emission region is 
not a slab, but a cylinder-like, a fraction of the black body radiation can 
reach the observer directly. This would reduce then the ``apparent'' optical 
depth $\tau$ and remove the inconsistency.
If the radiation pattern has a more complicated shape than that assumed here, 
the constraints on inclination and other model parameters become less strict. 

\begin{figure}
\bc
\psfig{figure=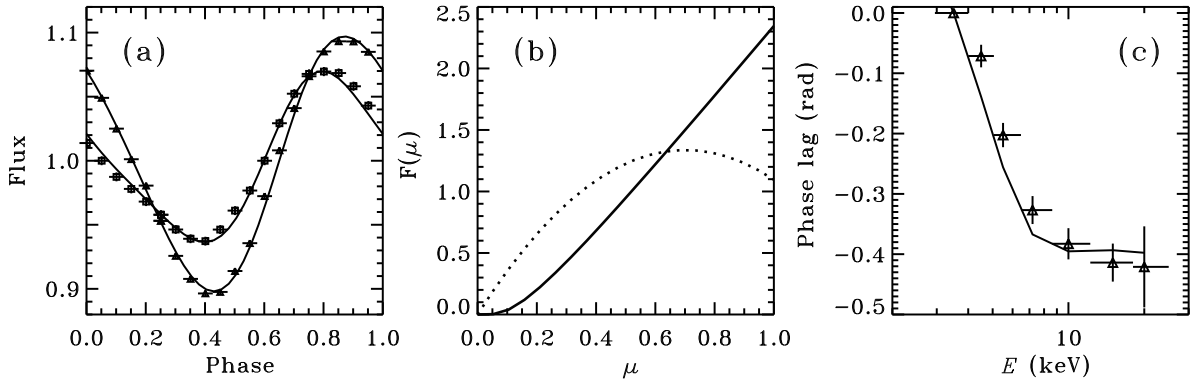,width=6.2in}
\ec
\caption{(a) The observed pulse profiles in the energy bands
3--4 keV (triangles) and 12--18 keV (squares) and the best-fit model light curves as a function of phase.
A neutron star with mass $M=1.6M_\odot$ and
radius $R=2 \rg$, and the inclination $i=75\deg$ are assumed.
The best-fit parameters are $\delta=12\fdg 2$, $\tau=0.22$, $b=-0.71$.
(b) The angular distribution of the black body and Comptonized fluxes
(normalized as $\int F(\mu)d \mu=1$).
(c) The observed (triangles) and the model (curve) phase lags  at the pulsar frequency
relative to the 3--4 keV band.
\label{fig:all}}
\end{figure}

One of the fits to the pulse profiles with the fixed $M$, $i$, and $R$ is 
shown in Fig.~\ref{fig:all}a. 
We reproduce easily the observed energy dependence of the pulse profiles. 
This model produces the phase lags at the pulsar frequency between different energies
similar to that observed (Fig.~\ref{fig:all}c). The corresponding angular 
distribution of the radiation flux is shown in Fig.~\ref{fig:all}b.

\section{Summary}

The two main spectral components (black body and a Comptonized) are clearly identified in the 
time-averaged spectrum of \sax.  The pulse profiles show significant energy dependence and 
the softer photons are lagging the harder ones. 
Such a behaviour can be easily reproduced by a model where hard Comptonized 
emission peaks earlier than the black body emission (Gierli\'nski et al. 2002). 
Due to the high velocity of the emitting region, the Doppler effect strongly 
affects the shape of the profiles. 
A black body  component shows about 25\% variability amplitude and the 
Doppler effect (producing a $\sim$ 6\%  variability) does not change 
it much. We showed that  the Comptonized emission from a slab of 
Thomson optical depth of $\tau\sim 0.3$ has a broader angular 
distribution (a ``fan''-like) and therefore an intrinsically smaller 
variability amplitude. 
The  Doppler effect then shifts significantly the peak of the 
pulse at high energies where the Comptonized emission dominates 
producing soft lags.

By fitting the observed pulse profiles we 
obtain constraints on the neutron star radius ($R=7.5\pm1.0$ km)
and the inclination of the system $i>60\deg$. 
We show that the angle between the magnetic dipole and the rotational axis 
is $\sim 20\deg$.

%\section*{Acknowledgments}
%This research was not supported by any grants.

\end{document}